\documentclass[12pt]{iopart}

\usepackage[final]{graphicx}

\newcommand{\ket}[1]{|#1\rangle}
\usepackage{SIunits}

\begin{document}

\title[]{Quantitative analysis of quantum dot dynamics and emission spectra in cavity quantum electrodynamics}

\author{K.~H.~Madsen$^{1,2}$ and P.~Lodahl$^2$}
\address{$^1$DTU Fotonik, Department of Photonics Engineering, Technical University of Denmark, \O rsteds Plads 343, DK-2800 Kgs.\ Lyngby, Denmark}
\address{$^2$Niels Bohr Institute, University of Copenhagen, Blegdamsvej 17, DK-2100 Copenhagen, Denmark}
\ead{khmadsen@nbi.ku.dk and lodahl@nbi.ku.dk}
\begin{abstract}
We present detuning-dependent spectral and decay-rate measurements to study the difference between spectral and dynamical properties of single quantum dots embedded in micropillar and photonic-crystal cavities. For the micropillar cavity, the dynamics is well described by the dissipative Jaynes-Cummings model, while systematic deviations are observed for the emission spectra. The discrepancy for the spectra is attributed to coupling of other exciton lines to the cavity and interference of different propagation paths towards the detector of the fields emitted by the quantum dot. In contrast, quantitative information about the system can readily be extracted from the dynamical measurements. In the case of photonic crystal cavities we observe an anti crossing in the spectra when detuning a single quantum dot through resonance, which is the spectral signature of strong coupling. However, time-resolved measurements reveal that the actual coupling strength is significantly smaller than anticipated from the spectral measurements and that the quantum dot is rather weakly coupled to the cavity. We suggest that the observed Rabi splitting is due to cavity feeding by other quantum dots and/or multiexcition complexes giving rise to collective emission effects.
\end{abstract}

\maketitle

\section{Introduction}
Cavity quantum electrodynamics (CQED) provides a way of enhancing and controlling the light-matter interaction between a single emitter and a cavity field and has potential applications in the field of quantum-information processing. This field was pioneered for atomic systems where a number of founding experimental demonstrations were achieved~\cite{Raimond.RevModernPhysics.2001} , while scaling these experiments to larger networks required for quantum-information processing remains a major challenge. Semiconductor quantum dots (QDs) embedded in nanophotonic structures offer an alternative and promising platform that currently is in its infancy, but could potentially lead to scalable quantum-information processing on an optical chip~\cite{Obrien.NaturePhotonics.2009} by exploiting the vast potential of semiconductor technology. Significant progress has been made in the field, where both Purcell enhancement~\cite{Gerard.PRL.1998,Gevaux.APL.2006} and strong coupling~\cite{Yoshie.Nature.2004,Reithmaier.Nature.2004} between a single QD and a nanocavity have been reported, and most recently non-Markovian dynamics~\cite{Madsen.PRL.2011} as well as few-photon non-linearities~\cite{Reinhard.NaturePhotonics.2012} have been demonstrated.

At a first glance, QDs have many properties in common with atoms, e.g., their quantized energy levels give rise to single-photon emission. Looking in more details reveals a number of effects unique to QDs, e.g., the point-dipole approximation may break down due to the mesoscopic size of QDs~\cite{Andersen.NaturePhysics.2011} and intrinsic exchange-mediated spin-flip processes can couple various finestructure exciton levels~\cite{Johansen.PRB.2010}. The proper understanding of the latter has enabled the use of QDs as probes of the local optical density of states (LDOS)~\cite{Wang.PRL.2011} as well as complete control over the spin state of the exciton with a single picosecond laser pulse~\cite{Kodriano.PRB.2012}. In addition, the presence of the solid-state environment leads to new phenomena, such as the creation of a quasi continuum of exciton states due to the interaction of electrons in the QD with electrons in the wetting layer~\cite{Winger.PRL.2009} or the observation of a phonon-assisted Purcell effect in a cavity, where the QD decay is stimulated by the exchange of phonons with a reservoir~\cite{Hohenester.PRB.2009,Madsen.arXiv.2012}.
\begin{figure}
\includegraphics[width=150mm]{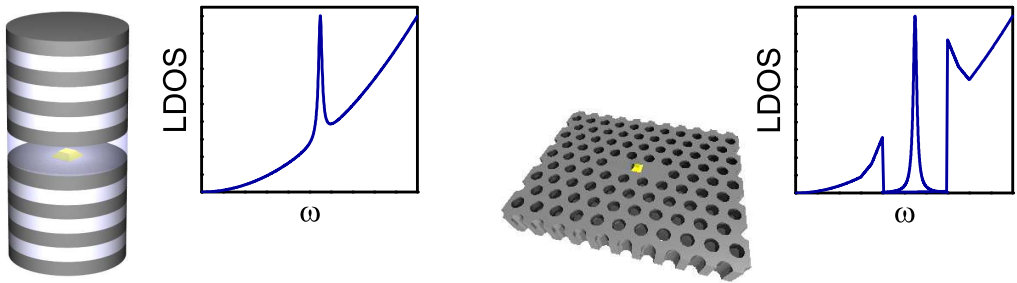}
\caption{Illustration of a QD (yellow point) in a micropillar cavity (left) and an L3 photonic crystal cavity (right) together with sketches of their respective LDOS.\label{fig:1}}
\end{figure}

In this paper, we present a comprehensive study of the dynamical and spectral properties of CQED systems including a quantitative comparison to theory. We perform experiments on single QDs embedded either in a micropillar cavity or in a photonic crystal (PC) cavity. In Fig.~\ref{fig:1} both physical systems are illustrated alongside sketches of the frequency variation of their respective LDOS. The central region of the micropillar cavity, where the QD is positioned, is surrounded by Bragg mirrors on each side giving rise to a symmetrically confined electric field perpendicular to the substrate surface. While the QD is randomly positioned radially, it is by design always situated at an antinode of the cavity field ensuring the effective coupling to the cavity. A sketch of the LDOS is shown in Fig.~\ref{fig:1}, where the sharp resonance at the cavity frequency reflects the build-up cavity field sitting on top of a background accounting for coupling to radiation modes. Figure~\ref{fig:1} also shows the PC cavity, where holes are periodically etched in a thin membrane with the QDs in the center and three holes on a row are not etched, thereby defining the PC cavity. The small size of the cavity allows for a tight confinement of light but the light-matter coupling strength is sensitive to the QD position relative to the antinode of the cavity field. The surrounding periodic structure gives rise to a 2D photonic band gap~\cite{Lodahl.Nature.2004}, which in Fig.~\ref{fig:1} is illustrated as the strong suppression of the LDOS in a wide frequency range, and the cavity resonance is inside the band gap.

In the following we show that while theory predicts well the dynamics of single QDs tuned in and out of resonance of both micropillar and PC cavities, this is not the case for the spectral measurements. This discrepancy is attributed to the fact that the details of the out coupling of the photons from the cavity is not well controlled in the experiment and mutual interference between different propagation paths is possible. For the PC cavity we observe an anti crossing in the measured spectra, when tuning the cavity through a QD resonance, which suggests strong coupling between the QD and the cavity. We determine the light-matter coupling strength from the observed Rabi splitting and also from time-resolved measurements of the decay rate and they differ by more than a factor of $3$. Thus the decay rate is found to be significantly slower than expected from the avoided crossing, which proves that the QD is in fact not in the strong-coupling regime despite the expectations from the spectral measurements. We attribute this difference to feeding of the cavity by other QD and multiexcitons that may give rise to a collective Rabi splitting.

\section{Theory}\label{Theory}
The interaction between a single emitter and a cavity mode can be described with the dissipative Jaynes-Cummings (JC) model~\cite{Carmichael.PRA.1989}. The QD is assumed to be a two-level system with an excited state, $\ket{e}$, and ground state, $\ket{g}$ that are coupled through the transition dipole moment, $\bf{d}_{\mathrm{eg}}$. The cavity field can either be in a single-photon state $\ket{1}$ or in the vacuum state $\ket{0}$ since we consider the case with at maximum one excitation in the system. The Hamiltonian describes the interaction between the emitter and the electric cavity field and is given by $\hat{H}=\hbar \omega_{qd} \hat{\sigma}_+ \hat{\sigma}_- +  \hbar \omega_{ca}\hat{a}^\dag \hat{a} + i\hbar g(\hat{\sigma}_- \hat{a}^\dag -\hat{\sigma}_+ \hat{a})$, where $g$ is the light-matter interaction strength, $\hat{\sigma}_-$ and $\hat{\sigma}_+$ ($\hat{a}$ and $\hat{a}^\dag$) are the annihilation and creation operators for the emitter (cavity field), and $\omega_{qd}$ ($\omega_{ca}$) is the QD (cavity field) frequency. The master equation is
\begin{eqnarray}\label{MasterEq}
\frac{\partial}{\partial t}\hat{\rho}=-\frac{i}{\hbar}\left[ \hat{H},\hat{\rho}\right] +\mathcal{L}(\gamma,\hat{\sigma}_-)+\mathcal{L}(\kappa,\hat{a})+\mathcal{L}(\gamma_{dp},\hat{\sigma}_z),
\end{eqnarray}
where $\hat{\rho}$ is the density matrix for the system. The first term in Eq.~(\ref{MasterEq}) expresses the coherent light-matter interaction, while the latter terms account for dissipation. Dissipative processes are included using Lindblad terms, $\mathcal{L}(\alpha,\hat{\sigma}) = \frac{\alpha}{2} \left(2 \hat{\sigma} \hat{\rho} \hat{\sigma}^\dag - \{ \hat{\sigma}^\dag \hat{\sigma} , \hat{\rho} \} \right)$, where the dissipation acting on operator $\hat{\sigma}$ has an associated rate of $\alpha$~\cite{Lindblad.Com.1976}. The decay of the emitter into leaky optical modes is described by the annihilation of an electronic excitation ($\hat{\sigma}_-$) with a rate $\gamma$, while leakage out of the cavity mode is described by the annihilation of a photon in the cavity field ($\hat{a}$) with the rate $\kappa$ due to the finite Q-factor, $Q=\omega_{ca}/\kappa$. Finally, decoherence from the solid-state environment (primarily due to phonons) is included as pure dephasing of the transition by the operator $\hat{\sigma}_z=[\hat{\sigma}_+,\hat{\sigma}_-]$ with the dephasing rate $\gamma_{dp}$.

From Eq.~(\ref{MasterEq}) we obtain the equations governing the dynamics of the system
\begin{eqnarray}\nonumber
\dot{\rho}_{qd}&=&-g(\rho_{po}+\rho^*_{po})-\gamma \rho_{qd}, \\
\dot{\rho}_{ca}&=&g(\rho_{po}+\rho^*_{po})-\kappa \rho_{ca},\\
\dot{\rho}_{po}&=&g(\rho_{qd}-\rho_{ca})-\left(\gamma_{tot}+i\Delta\right) \rho_{po}, \nonumber
\end{eqnarray}
where $\gamma_{tot}=(\kappa+\gamma+2\gamma_{dp})/2$, $\Delta=\omega_{qd}-\omega_{ca}$, $\rho_{qd}$ ($\rho_{ca}$) is the population of the emitter (cavity mode), and $\rho_{po}$ is proportional to the polarization. In the weak-coupling regime the QD decays irreversibly and the polarization can be adiabatically eliminated by setting $\dot{\rho}_{po}=0$, where the decay rate of the QD becomes $\Gamma=\gamma + 2g^2 \frac{\gamma_{tot}}{\gamma_{tot}^2+\Delta^2}$. When the light-matter coupling rate is sufficiently large compared to the dissipation, the cavity system enters the strong-coupling regime where a photon is stored so long in the cavity that it can be reabsorbed by the emitter. As a result the population of the emitter undergoes Rabi oscillations. The spectral signature of strong coupling is the anti crossing of the QD and cavity peak when tuning them into mutual resonance.

Although the QD is often treated as a two-level emitter, the actual electronic structure is more complicated. The QD excitons predominately recombine radiatively due to the high quantum efficiency, but non-radiative processes do occur as well~\cite{Johansen.PRB.2008}. Furthermore, additional exciton states that cannot recombine radiatively (dark-exciton states) are also populated in the QD and can couple to the radiative states (bright-exciton states) through a slow spin-flip process~\cite{Johansen.PRB.2010}. As a consequence, the time-resolved emission from a single QD under non-resonant excitation is bi-exponential, where the fast decay rate corresponds to the decay of the bright state, which is the relevant rate in the present experiments. Finally, an electron confined in the QD can also scatter on electrons in the wetting layer, thereby giving rise to a quasi continuum of multiexcitonic states~\cite{Winger.PRL.2009}. These states are found to be responsible for the pronounced QD-cavity coupling observed in spectral measurements even for very large detunings~\cite{Hennessy.Nature.2007,Ates.NaturePhotonics.2009}.

The total emission spectrum is obtained by the Wiener-Khinchin theorem~\cite{Loudon} according to
\begin{eqnarray}\label{spectra_Raymer}
S_{det}(\omega)=\frac{2}{\pi} Re \left[ \int_0^\infty d\tau e^{i(\omega-\omega_{qd})
\tau} \int_0^\infty dt' \langle \hat{E}_{det}^{(-)}(t'+\tau)\hat{E}_{det}^{(+)}(t') \rangle \right],
\end{eqnarray}
where $\hat{E}_{det}^{(+)}$ ($\hat{E}_{det}^{(-)}$) is the positive (negative) frequency part of the electric field that reaches the detector. This field is related to the emitter and cavity field operator through, $\hat{E}^{(+)}_{det}(t) =\eta_{ca}\sqrt{\kappa}\hat{a}(t)+\eta_{qd}
\sqrt{\gamma}\hat{\sigma}_-(t)$, where we neglect any time retardation between the emitter and cavity operators, which is usually a good approximation in nanophotonics cavities. The coefficients $\eta_{ca}$ and $\eta_{qd}$ are complex coefficients that describe the collection efficiencies of the cavity and QD electric field, respectively, and their relative phase. The quantum regression theorem in differential form~\cite{Perea.PRB.2004} states that the two-time expectation values in Eq.~(\ref{spectra_Raymer}) (e.g., $\langle \hat{a}^\dag (t'+\tau) \hat{a} (t') \rangle$) has the same time evolution as the corresponding one-time expectation value ($\langle \hat{a}^\dag (t'+\tau) \rangle$). We can then express the total emitted spectrum as
\begin{eqnarray}\label{full_spectrum}
S_{det}(\omega)=Re\left( |\eta_{ca}|^2 S_{ca}+|\eta_{qd}|^2 S_{qd}+\eta_{ca}\eta^*_{qd} S_{qd,ca}+\eta^*_{ca}\eta_{qd} S^*_{qd,ca}\right),
\end{eqnarray}
where the two first terms are the cavity and QD spectra while the latter two terms account for the interference between the cavity and QD electric fields. These interference terms are usually neglected in the literature, but are expected to play a pronounced role if the details of the emission spectra should be reproduced. In Fig.~\ref{fig:2} a micropillar cavity is illustrated, and the cavity and QD spectra are drawn schematically to indicate that the cavity field is expected to predominately leak out of the top of the micropillar, while the QD field should leak predominantly in the radial direction. Nonetheless in an experiment it is not likely that the two contributions can be fully separated leading to interference, and in PC cavities the interference is expected to be even more pronounced.

\section{Micropillar Cavities}
\begin{figure}
\includegraphics[width=150mm]{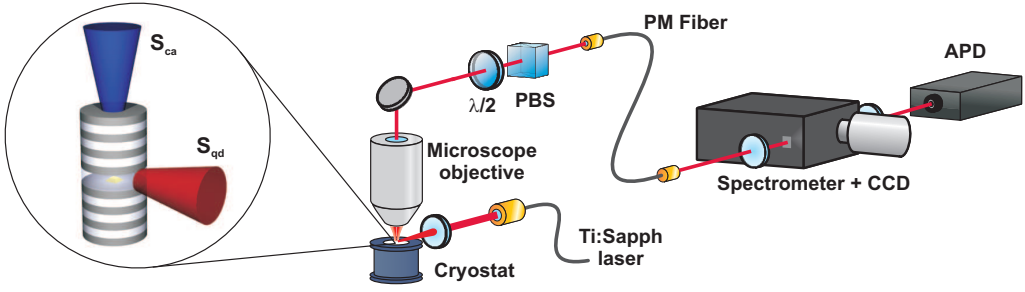}
\caption{Experimental setup used for measuring spectra and decay curves. The micropillar or PC cavity is placed in a He flow cryostat capable of cooling down to temperatures of $4.2$ K. The micropillar is optically excited under a $15$ degree angle relative to the substrate surface. The emission is collected by an objective (N.A.=0.6), and after polarization and spatial selection the emission is spectrally resolved by a spectrometer and sent to a CCD (APD) for spectral (dynamical) measurements. For measurements on the PC cavity excitation is vertical and a dichroic mirror is inserted above the microscope for separating the excitation laser from the emission. On the left a sketch of the micropillar cavity is shown together with illustrations of the primary leak directions for the cavity field (spectrum $S_{ca}$) and QD field (spectrum $S_{qd}$).\label{fig:2}}
\end{figure}
The micropillar cavity has a height of $\sim 9\; \micro \metre$, a diameter of $1.7 \; \micro\metre,$ and consists of alternating GaAs and AlAs layers surrounding a central GaAs cavity that contains a low density of self-assembled InAs QDs ($60-90$ \micro \metre$^{-2}$)~\cite{Loffler.APL.2005}. The sample is placed in a He flow cryostat and optically excited from the side with a beam that has a $15$ degrees angle of incidence relative to the substrate surface, cf. Fig.~\ref{fig:2}, which enables an efficient separation of the emission from the sample and the excitation beam. The emission from the micropillar is collected using a microscope objective (N.A. = $0.6$), and with a half-wave plate ($\lambda/2$) and a polarizing beam splitter (PBS) a single polarization component is selected. Spatial filtering is performed by coupling into a single mode polarization maintaining (PM) fiber, and a spectrometer resolves the frequency components of the emission. In the spectral measurements, the emitted light is imaged onto a CCD camera, while for time-resolved measurements it is directed through a slit that singles out a narrow frequency band and on to an avalanche photodiode (APD). The QD is excited using $3$ ps long pulses from a Ti:Sapph. laser and in order to selectively excite only the QD of interest we tune the excitation wavelength of the laser into resonance with the p-shell of the QD. This enables us to strongly suppress the emission from other QDs and autocorrelation measurements verify that we detect emission from a single QD.

\begin{figure}
\includegraphics[width=150mm]{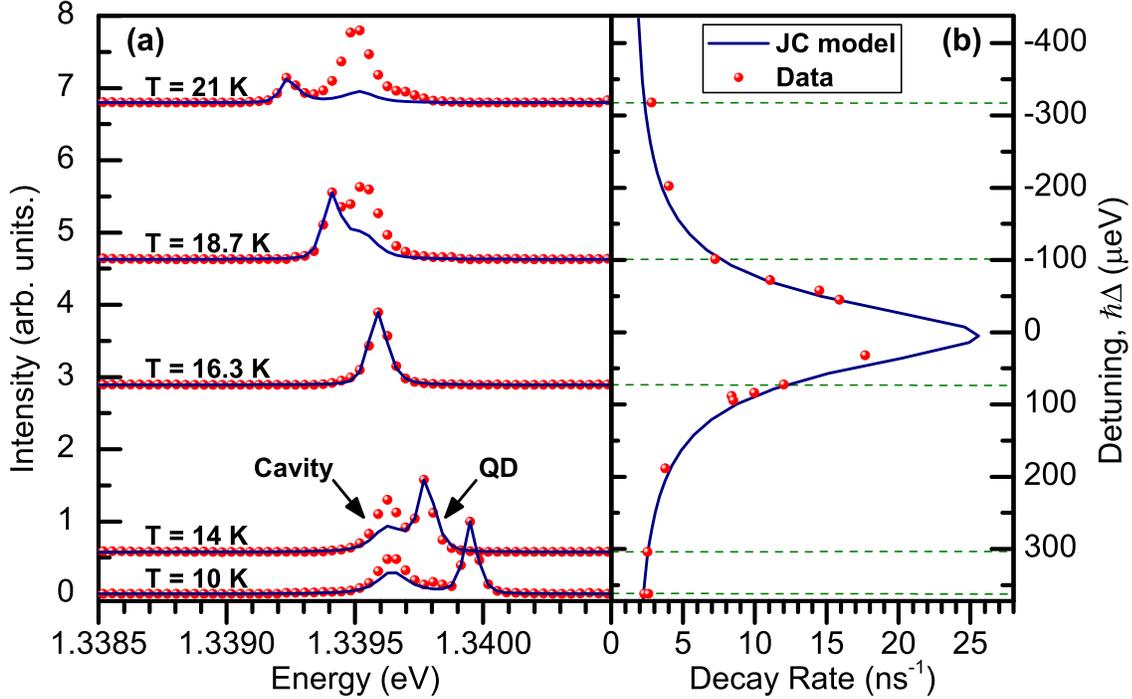}
\caption{\textbf{a)} The measured spectra (red circles) for a QD in a micropillar cavity for various values of the detuning that is controlled through the temperature $T$. The blue line is the corresponding theory without any free fitting parameters, as explained in the text. \textbf{b)} Measured mean decay rate (red circles) as a function of detuning with a comparison to the theoretical predictions (blue line). These decay rate data have been published previously in Ref.~\cite{Madsen.PRL.2011} .\label{fig:3}}
\end{figure}

We systematically vary the detuning by controlling temperature, and both the spectrum and decay curve of the QD emission are recorded versus detuning. In Fig.~\ref{fig:3}(a) the spectra for a few selected detunings are presented, and a clear crossing of the QD and cavity mode is observed at resonance indicating that the cavity is not in the strong-coupling regime. We note that the cavity intensity relative to the QD intensity is strongly asymmetric with respect to detuning. The Q-factor can be extracted from spectral measurements by using strong above-band excitation to ensure that the QDs are saturated and the cavity spectrum is recorded. After deconvoluting with the spectrometer instrument-response function (IRF) we find $Q=12200$ corresponding to $\hbar\kappa=110 \;\micro$eV. In Fig.~\ref{fig:3}(b) we present the mean decay rate versus detuning obtained from the measured decay curves~\cite{Madsen.PRL.2011}, where we use a spectrometer to filter out a narrow region around the QD frequency. In the present experiment the cavity is in an intermediate-coupling regime close to strong coupling where the dynamics deviates from the exponential decay, which is the signature of non-Markovian coupling to a radiative reservoir. As a consequence we extract the mean decay rate directly from the decay curves, which is the inverse of the of the mean decay time. A very pronounced Purcell enhancement within a frequency range that is limited by the Q of the cavity is displayed in Fig.~\ref{fig:3}(b). The data can be compared quantitatively to theory without any adjustable parameters after independently determining all the governing parameters: From decay-rate measurements at a large detuning the QD decay rate to leaky modes is determined, $\hbar\gamma=1.3\; \micro$eV and a Hong-Ou-Mandel interferometer~\cite{Santori.Nature.2002} enables determining a pure dephasing rate of $\hbar\gamma_{dp}=6.3 \; \micro$eV at $\mathrm{T}=16.3$ K. The fastest observed average decay rate is $\Gamma=17.7$ ns$^{-1}$ at $\hbar\Delta=17\; \micro$eV, which in combination with the values for $\gamma$, $\kappa$, and $\gamma_{dp}$, allows us to extract the coupling strength $\hbar g=22.6\;\micro\mathrm{eV}$. Using these parameters the calculated mean decay rate is observed to be in excellent agreement with the experimental data, cf. Fig.~\ref{fig:3}(b), clearly illustrating that a complete quantitative understanding of the dynamics is found without the need for any adjustable parameters.

We now perform a similar comparison between experiment and theory for the spectral measurements. As explained earlier, the full spectrum in Eq.~(\ref{full_spectrum}) has several contributions, but the individual prefactors are undetermined and we will assume $\eta_{qd}=0$ since the cavity spectrum is expected to be dominating in the micropillar geometry. At resonance ($\Delta=0$), we also perform a Hanbury Brown-Twiss measurement and record an autocorrelation of $g^{(2)}(0)=34.5\%$, which indicates additional feeding of the cavity from other QDs and multiexcitons~\cite{Winger.PRL.2009}. As a result, the background contribution to the cavity spectrum will by assuming a thermal distribution constitute the fraction $\frac{g^{(2)}(0)}{2-g^{(2)}(0)}$, which in our values can be evaluated to be $20.8\%$. We include this contribution in the calculated spectra by adding an inhomogeneous background of $20.8\%$ to the cavity. In Fig.~\ref{fig:3}(a) the calculated spectra are shown where the experimental parameters from the time-resolved measurements are employed and each spectrum has been convoluted with the measured IRF and normalized to the QD peak. We observe that the calculated spectra consistently underestimate the cavity intensity. This disagreement pinpoints that the asymmetry in the measured relative QD-cavity intensity cannot be reproduced by the theory. Indeed a number of effects are not included in the theory that do influence spectral measurements as opposed to the dynamical measurements.  Thus, the inability to include the additional interference terms between the QD and cavity emissions in Eq.~(\ref{full_spectrum}) will influence the comparison in particular because the  coefficients $\eta_{ca}$ and $\eta_{qd}$ are likely to be detuning dependent. We note that even if we include these coefficients as free parameters, the agreement between calculated and measured spectra is still very poor. Furthermore, feeding from other excitons is known to increase the intensity of the cavity even when they are detuned far away. Finally, the time-resolved measurements, including Hanbury Brown-Twiss and Hong-Ou-Mandel measurements, probe the dynamics and coherence at short timescales ($\sim\;$ns) while the spectra are integrated over a much longer timescale ($\sim\;$s). This implies that spectral measurements are sensitive to slow dephasing mechanisms, e.g., spectral diffusion, which could give rise to additional broadening compared to dynamical measurements. These complications lead to the conclusion that quantitative information about the QD-based cavity QED system is favorably extracted from time-resolved measurements rather than in the spectral domain, where systematic deviations between experiment and theory are generally observed.

\section{Photonic Crystal Cavities}
For comparison, we have performed a similar study for a single QD tuned through a PC cavity. We investigate a GaAs PC membrane with lattice constant $a=240 \; \mathrm{nm}$, hole radius $r=65\; \mathrm{nm}$, and a width of $154$ nm, where a layer of self-assembled InAs QDs has been grown in the middle of the membrane with a density of $\sim 80 \; \micro \metre^{-2}$. We introduce an L3 cavity by leaving out three holes, and in order to increase the Q-factor, the first three holes at each end of the cavity are shifted by $0.175a$, $0.025a$, and $0.175a$, respectively~\cite{Akahane.OpticsExpress.2005}. The sample is characterized in the same experimental setup (see Fig.~\ref{fig:2}), except that the cavity is excited from the top and a dichroic mirror after the microscope objective separates the excitation laser from the emission. The pulsed excitation laser is tuned into resonance with a higher order mode (M6) of the cavity at $850$ nm, while we observe the fundamental high-Q mode (M1) at $952$ nm. This excitation scheme allows us to selectively excite QDs that are spatially coupled to the cavity. The QD emission frequency redshifts with increasing temperature, while the cavity frequency redshifts when small amounts of nitrogen is deposited on the sample. Utilizing these two techniques in combination gives a way to control the detuning between the QD and the cavity mode over a large frequency range. We record $Q=6690$ (equal to $\hbar\kappa=195 \; \micro$eV) by pumping the QDs into saturation similarly to the measurements described for the micropillar cavity.

\begin{figure}
\includegraphics{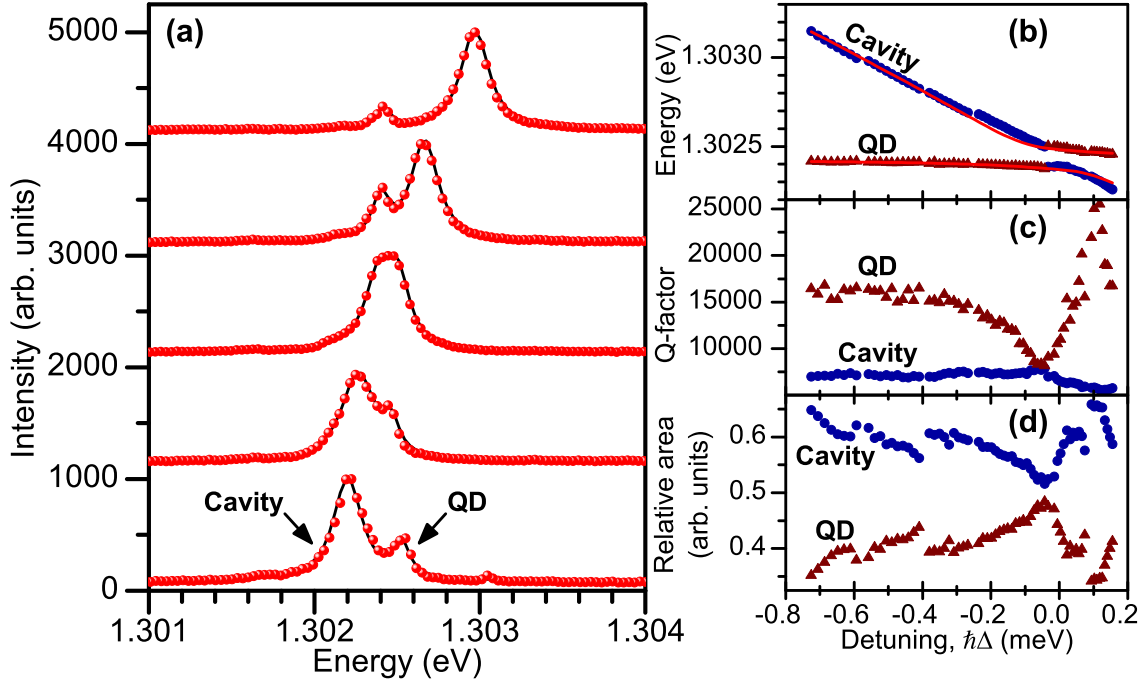}
\caption{\textbf{a)} Measured emission spectra (red circles) of a QD tuned through resonance of a PC cavity by deposition of N$_2$ on the sample while the temperature is kept constant at T $=10$ K. The black line is the fit with two lorentzians after convolution with the spectrometer IRF. \textbf{(b-d)} The resonance energy, Q-factor, and relative area of the two peaks, respectively, as a function of detuning. Red lines are the fit of the cavity spectra from the JC model used to extract the coupling strength.
The relative area of, e.g., the QD is defined as A$_{\mathrm{qd}}/$(A$_{\mathrm{qd}}$+A$_{\mathrm{ca}}$), where A$_{\mathrm{qd}}$ (A$_{\mathrm{ca}}$) is the QD (cavity) area.
\label{fig:4}}
\end{figure}
 Figure~\ref{fig:4}(a) shows emission spectra for different values of detuning where the cavity is tuned through the QD while the temperature is fixed at T$=10$ K. The measured spectra are deconvoluted by performing an inverse Fourier transform, divide by the deconvoluted IRF, and bandpass filtering in order to reduce noise, before finally Fourier transforming it back into frequency space. The validity of this procedure is carefully checked by convoluting the deconvoluted spectra with the IRF and comparing it to the measured spectra. The experimental data can be fitted well by the sum of two lorentzians, see  Fig.~\ref{fig:4}(a). We stress that in order to obtain a successful fit, the center, width, and the heights of the lorentzians are free parameters. The limitations to the quantitative knowledge that can be extracted from the spectra, as was discussed for the micropillar cavities above, apply also to the case of PC cavities. Here we will focus on investigating the Rabi splitting that has been widely studied in the literature~\cite{Yoshie.Nature.2004,Reithmaier.Nature.2004,Hennessy.Nature.2007} and believed to be a robust measure of the coupling of the cavity system.

Figure~\ref{fig:4}(b-d) show the quantities extracted from modeling the spectra, i.e., the energy of the two resonances, the associated Q-factors, and the relative area of each of the peaks. From the peak energies (Fig.~\ref{fig:4}(b)), we observe the anti-crossing of the cavity and QD peak when tuned into mutual resonance, which is the spectral signature of strong coupling. Furthermore, we observe in Fig.~\ref{fig:4}(c) that the QD linewidth broadens (Q-factor decreases) and equals the cavity linewidth, which is again a feature found in the strong-coupling regime where two indistinguishable alternatives exist whether the photon is emitted to the cavity or absorbed by the QD.
For large negative detunings, the Q of the QD resonance is constantly around $\sim 15000$, while it rises to about $\sim25000$ for positive detuning, which is likely an effect of the deconvolution process since the QD linewidth is so narrow that it becomes comparable to the resolution of the spectrometer (Q$=43500$). The cavity Q-factor does not change significantly over the detuning range. Close to resonance, however, a small increase is observed, but we note that for a strongly-coupled cavity the Q-factor is expected to almost double. Finally, the relative areas of the lorentzians are plotted in Fig.~\ref{fig:4}(d), and we observe that the cavity peak dominates the spectrum apart from close to resonance, where the areas become similar.

\begin{figure}
\includegraphics{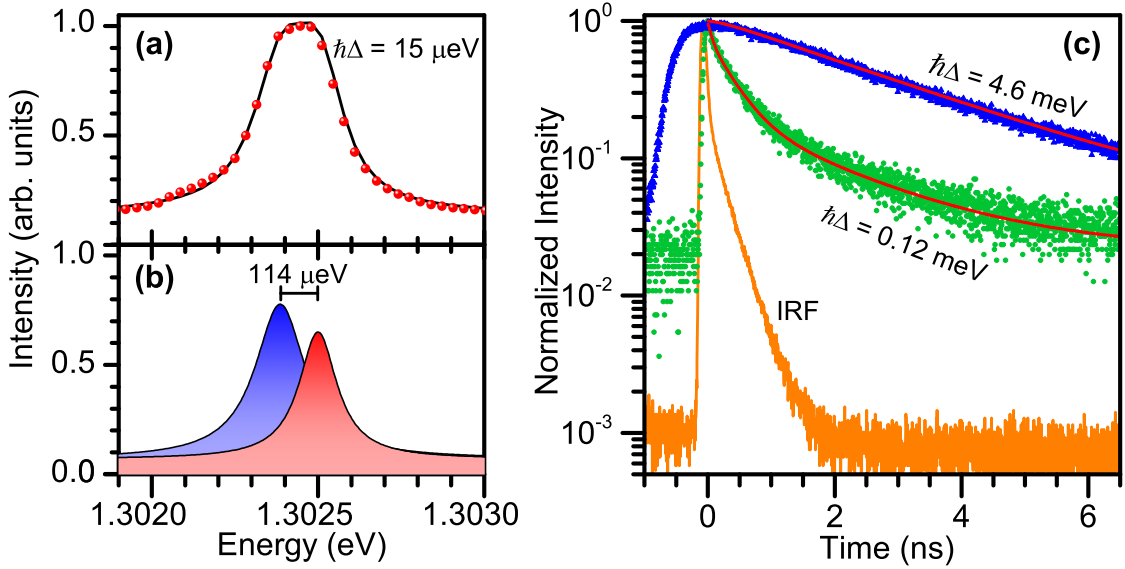}
\caption{\textbf{a)} Measured emission spectrum (red circles) at a detuning of $\hbar\Delta=15\;\micro$eV together with the double lorentzian fit (black line). \textbf{b)} The deconvoluted double lorentzian function revealing a Rabi splitting of $\hbar\Omega=114\;\micro$eV. \textbf{c)} Measured decay curves close to resonance ($\hbar\Delta=0.12$ meV) and far detuned ($\hbar\Delta=4.6$ meV) together with the fits of the decay curves (red lines). We extract the decay rates $18.5 \; \mathrm{ns}^{-1}$ and $0.39 \; \mathrm{ns}^{-1}$, respectively. For reference the IRF of the APD is shown.\label{fig:5}}
\end{figure}

Figure~\ref{fig:5}(a) displays the emission spectrum recorded almost at resonance $(\hbar \Delta=15 \; \micro$eV) together with the double lorentzian fit, and in Fig.~\ref{fig:5}(b) the fit is decomposed into the two lorentzians. The two lorentzians have almost same width and area, as also can be seen from the data in  Fig.~\ref{fig:4}(c-d), which is expected for a system in the strong-coupling regime. The splitting between the two peaks is $\hbar \Omega=114 \; \micro$eV. Assuming that the splitting originates from a single QD strongly coupled to the cavity, we fit the cavity spectrum from the JC model~\cite{Laussy.PRB.2011} to the data using the experimentally measured values of $\kappa$ and $\gamma$, and a dephasing rate of $\hbar \gamma_{dp}=4 \; \micro$eV~\cite{Madsen.PRL.2011}. The result is shown in Fig.~\ref{fig:4}(b) and we extract a coupling strength of $\hbar g=92.4 \; \micro$eV.

As was described in the section on micropillar cavities, a quantitative measure of the coupling strength can be obtained from time-resolved measurements thereby testing the validity of the observed Rabi splitting. Figure~\ref{fig:5}(c) shows examples of decay curves of the QD recorded close to and far from resonance, respectively. A strong Purcell enhancement of close to $50$ is observed for the fast decay rate of the recorded decay curve. We note that for detunings $\hbar |\Delta|<0.5$ meV additional exciton lines feed the cavity and the decay curves are multiexponential, and we extract the fast rate that will be dominated by the resonant exciton. Figure~\ref{fig:5}(c) displays also the IRF of the APD that is taken into account by convoluting the model with the IRF before fitting to the measured decay curves. For the largest detuning we observe a decay rate of $0.39 \; \mathrm{ns}^{-1}$ and here the coupling to the cavity is negligible so that the background decay rate associated with coupling to leaky modes is $\hbar \gamma=0.2 \; \micro$eV. The detuning-dependent decay rates have been studied in detail in \cite{Madsen.arXiv.2012}, where comparison between experiment and theory enabled determining the effective phonon density of states. Close to resonance ($\hbar \Delta=120\; \micro$eV) a decay rate as large as $18.5 \; \mathrm{ns}^{-1}$ is observed. We use the expression for the Purcell factor~\cite{Gerard.TopicsAP.2003} to conclude that the coupling strength in the PC cavity is only $3.2\%$ of the maximum achievable value for an emitter positioned optimally in an antinode of the cavity with aligned dipole moment. For comparison this value is $19\%$ for the reported experiment in micropillars. The pronounced deviations from ideal coupling could be due to unavoidable imperfections in the nanophotonic cavities that may influence the local coupling strength, and a systematic study of disorder in PCs has been reported in~\cite{Garcia.arxiv.2012}. From the data recorded close to resonance we determine the coupling strength of $\hbar g=22\; \micro$eV. Surprisingly, the coupling strength determined from the dynamic measurements is found to be less than one fourth of the value obtained from the observed Rabi splitting and in fact proves that the cavity is in the weak rather than the strong-coupling regime. This pronounced discrepancy is another example of the incompatibility of the information extracted from spectral and dynamical measurements. In the time-resolved measurements only a narrow spectral region is selected around the QD line while in the spectral measurements also the influence of other QDs and multi-exciton transitions feeding the cavity are observed~\cite{Ates.NaturePhotonics.2009,Kaniber.PRB.2008}. This has been confirmed experimentally from measurements of the autocorrelation of the cavity peak where bunching has been observed despite the fact that the cavity was primarily fed by a single QD~\cite{Winger.PRL.2009}. Thus we suggest that the additional feeding of the cavity gives rise to collective coupling to the cavity that can significantly increase the Rabi splitting as predicted theoretically~\cite{Diniz.PRA.2011}. This mechanism could also potentially explain the surprisingly large Rabi splitting observed in other strong-coupling experiments that initially were suggested to be due to a giant oscillator strength of the large QDs~\cite{Reithmaier.Nature.2004}, but this explanation was found to be insufficient in detailed measurements of the oscillator strength~\cite{Stobbe.PRB.2010}.

\section{Conclusion}
We have performed a quantitative comparison between spectral and dynamical measurements of solid-state cavity QED systems. The dynamics of a single QD in a micropillar cavity is well described by the JC model, while using the same theory and parameters to compute the spectra reveals a large disagreement. The emission spectrum contains both a cavity and a QD part as well as interference terms between them, and the lack of detailed microscopic insight into the parameters determining their mixing limits the quantitative modeling of the experimental data. We have also presented time-resolved measurements of a QD in a PC cavity and observed pronounced Purcell enhancement enabling extracting the light-matter coupling strength. From this analysis the cavity is found to be in the weak-coupling regime, but nonetheless spectral measurements reveal a clear anti crossing with a pronounced Rabi splitting. The  observed Rabi splitting is likely a consequence of cavity feeding from other QDs and multiexciton complexes that induce a collective Rabi splitting. Such collective coupling of more QDs to the same cavity mode has been proposed as a way of creating a long-lived solid-state quantum memory~\cite{Diniz.PRA.2011}.

We thank S.~Ates and A.~Kreiner-M{\o}ller for assistance on the experiments and A. L\"{o}ffler, S. Reitzenstein, and A. Forchel for the collaboration on micropillar cavities that has lead to the data in Fig. (\ref{fig:3}) that were published in Ref. \cite{Madsen.PRL.2011}. We gratefully acknowledge  financial support from the Villum Foundation,
the Danish Council for Independent Research (Natural Sciences and Technology and Production
Sciences) and the European Research Council (ERC consolidator grant "ALLQUANTUM").
\section*{References}
\bibliographystyle{unsrt}
\bibliography{mybib}

\begin{thebibliography}{10}

\bibitem{Raimond.RevModernPhysics.2001}
J.~M. Raimond, M.~Brune, and S.~Haroche.
\newblock Colloquium: Manipulating quantum entanglement with atoms and photons
  in a cavity.
\newblock {\em Rev. Mod. Phys.}, 73:565--582, 2001.

\bibitem{Obrien.NaturePhotonics.2009}
J.~L. O'Brien, A.~Furusawa, and J.~Vuckovic.
\newblock Photonic quantum technologies.
\newblock {\em Nature Photoncis}, 3:687--695, 2009.

\bibitem{Gerard.PRL.1998}
J.-M. G\'{e}rard, B.~Sermage, B.~Gayral, B.~Legrand, E.~Costard, and
  V.~Thierry-Mieg.
\newblock Enhanced spontaneous emission by quantum boxes in a monolithic
  optical microcavity.
\newblock {\em Phys. Rev. Lett.}, 81:1110--1113, 1998.

\bibitem{Gevaux.APL.2006}
D.~G. Gevaux, A.~J. Bennett, R.~M. Stevenson, A.~J. Shields, P.~Atkinson,
  J.~Griffiths, D.~Anderson, G.~A.~C. Jones, and D.~A. Ritchie.
\newblock Enhancement and suppression of spontaneous emission by temperature
  tuning inas quantum dots to photonic crystal cavities.
\newblock {\em Appl. Phys. Lett.}, 88:131101, 2006.

\bibitem{Yoshie.Nature.2004}
T.~Yoshie, A.~Scherer, J.~Hendrickson, G.~Khitrova, H.~M. Gibbs, G.~Rupper,
  C.~Ell, O.~B. Shchekin, and D.~G. Deppe.
\newblock Vacuum rabi splitting with a single quantum dot in a photonic crystal
  nanocavity.
\newblock {\em Nature}, 432:200--203, 2004.

\bibitem{Reithmaier.Nature.2004}
J.~P. Reithmaier, G.~Sek, A.~L\"{o}ffler, C.~Hofmann, S.~Kuhn, S.~Reitzenstein,
  L.~V. Keldysh, V.~D. Kulakovskii, T.~L. Reinecke, and A.~Forchel.
\newblock Strong coupling in a single quantum dot-semiconductor microcavity
  system.
\newblock {\em Nature}, 432:197--200, 2004.

\bibitem{Madsen.PRL.2011}
K.~H. Madsen, S.~Ates, T.~Lund-Hansen, A.~L\"{o}ffler, S.~Reitzenstein,
  A.~Forchel, and P.~Lodahl.
\newblock Observation of non-markovian dynamics of a single quantum dot in a
  micropillar cavity.
\newblock {\em Phys. Rev. Lett.}, 106:233601, 2011.

\bibitem{Reinhard.NaturePhotonics.2012}
A.~Reinhard, T.~Volz, M.~Winger, A.~Badolato, K.~J. Hennessy, E.~L. Hu, and
  A.~Imamoglu.
\newblock Strongly correlated photons on a chip.
\newblock {\em Nature Photonics}, 6:93--96, 2012.

\bibitem{Andersen.NaturePhysics.2011}
M.~L. Andersen, S.~Stobbe, A.~S. S{\o}rensen, and P.~Lodahl.
\newblock Strongly modified plasmon-matter interaction with mesoscopic quantum
  emitters.
\newblock {\em Nature Physics}, 7:215--218, 2011.

\bibitem{Johansen.PRB.2010}
J.~Johansen, B.~Julsgaard, S.~Stobbe, J.~M.. Hvam, and P.~Lodahl.
\newblock Probing long-lived dark excitons in self-assembled quantum dots.
\newblock {\em Phys. Rev. B}, 81:081304, 2010.

\bibitem{Wang.PRL.2011}
Q.~Wang, S.~Stobbe, and P.~Lodahl.
\newblock Mapping the local density of optical states of a photonic crystal
  with single quantum dots.
\newblock {\em Phys. Rev. Lett.}, 107:167404, 2011.

\bibitem{Kodriano.PRB.2012}
Y.~Kodriano, I.~Schwartz, E.~Poem, Y.~Benny, R.~Presman, T.~A. Truong, P.~M.
  Petroff, and D.~Gershoni.
\newblock Complete control of a matter qubit using a single picosecond laser
  pulse.
\newblock {\em Phys. Rev. B}, 85:241304, 2012.

\bibitem{Winger.PRL.2009}
M.~Winger, T.~Volz, G.~Tarel, S.~Portolan, A.~Badolato, K.~J. Hennessy, E.~L.
  Hu, A.~Beveratos, J.~Finley, V.~Savona, and A.~Imamoglu.
\newblock Explanation of photon correlations in the far-off-resonance optical
  emission from a quantum-dot-cavity system.
\newblock {\em Phys. Rev. Lett.}, 103:207403, 2009.

\bibitem{Hohenester.PRB.2009}
U.~Hohenester, A.~Laucht, M.~Kaniber, N.~Hauke, A.~Neumann, A.~Mohtashami,
  M.~Seliger, M.~Bichler, and J.~J. Finley.
\newblock Phonon-assisted transitions from quantum dot excitons to cavity
  photons.
\newblock {\em Phys. Rev. B}, 80:201311, 2009.

\bibitem{Madsen.arXiv.2012}
K.~H. Madsen, P.~Kaer, A.~Kreiner-M{\o}ller, S.~Stobbe, A.~Nysteen, J.~M{\o}rk,
  and P.~Lodahl.
\newblock Measuring the effective phonon density of states of a quantum dot.
\newblock arXiv:1205.5623, 2012.

\bibitem{Lodahl.Nature.2004}
P.~Lodahl, A.~F. Driel, I.~S. Nikolaev, A.~Irman, K.~Overgaag,
  D.~Vanmaekelbergh, and W.~L. Vos.
\newblock Controlling the dynamics of spontaneous emission from quantum dots by
  photonic crystals.
\newblock {\em Nature}, 430:654--657, 2004.

\bibitem{Carmichael.PRA.1989}
H.~J. Carmichael, R.~J. Brecha, M.~G. Raizen, H.~J. Kimble, and P.~R. Rice.
\newblock Subnatural linewidth averaging for coupled atomic and cavity-mode
  oscillators.
\newblock {\em Phys. Rev. A}, 40:5516--5519, 1989.

\bibitem{Lindblad.Com.1976}
G.~Lindblad.
\newblock On the generators of quantum dynamical semigroups.
\newblock {\em Commun. math. Phys.}, 48:119--130, 1976.

\bibitem{Johansen.PRB.2008}
J.~Johansen, S.~Stobbe, I.~S. Nikolaev, T.~Lund-Hansen, P.~T. Kristensen,
  J.~M.. Hvam, W.~L. Vos, and P.~Lodahl.
\newblock Size dependence of the wavefunction of self-assembled inas quantum
  dots from time-resolved optical measurements.
\newblock {\em Phys. Rev. B}, 77:073303, 2008.

\bibitem{Hennessy.Nature.2007}
K.~Hennessy, A.~Badolato, M.~Winger, D.~Gerace, M.~Atat\"{u}re, S.~Gulde,
  S.~F\"{a}lt, E.~L. Hu, and A.~Imamoglu.
\newblock Quantum nature of a strongly coupled single quantum dot-cavity
  system.
\newblock {\em Nature}, 445:896--899, 2007.

\bibitem{Ates.NaturePhotonics.2009}
S.~Ates, S.~M. Ulrich, A.~Ulhaq, S.~Reitzenstein, A.~L\"{o}ffler,
  S.~H\"{o}fling, A.~Forchel, and P.~Michler.
\newblock Non-resonant dot-cavity coupling and its potential for resonant
  single-quantum-dot spectroscopy.
\newblock {\em Nature Photonics}, 3:724--728, 2009.

\bibitem{Loudon}
R.~Loudon.
\newblock {\em The Quantum Theory of Light}.
\newblock Oxford Science Publications, 3rd edition, 2000.

\bibitem{Perea.PRB.2004}
J.~I. Perea, D.~Porras, and C.~Tejedor.
\newblock Dynamics of the excitations of a quantum dot in a microcavity.
\newblock {\em Phys. Rev. B}, 70:115304, 2004.

\bibitem{Loffler.APL.2005}
A.~L\"{o}ffler, J.~P. Reithmaier, G.~Sek, C.~Hofmann, S.~Reitzenstein, M.~Kamp,
  and A.~Forchel.
\newblock Semiconductor quantum dot microcavity pillars with high-quality
  factors and enlarged dot dimensions.
\newblock {\em Appl. Phys. Lett.}, 86:111105, 2005.

\bibitem{Santori.Nature.2002}
C.~Santori, D.~Fattal, J.~Vuckovic, G.~S. Solomon, and Y.~Yamamoto.
\newblock Indistinguishable photons from a single-photon device.
\newblock {\em Nature}, 419:594--597, 2002.

\bibitem{Akahane.OpticsExpress.2005}
Y.~Akahane, T.~Asano, B.-S. Song, and S.~Noda.
\newblock Fine-tuned high-q photonic-crystal nanocavity.
\newblock {\em Optics Express}, 13:1202, 2005.

\bibitem{Laussy.PRB.2011}
F.~P. Laussy, A.~Laucht, E.~del Valle, J.~J. Finley, and J.~M. Villas-B\^{o}as.
\newblock Luminescence spectra of quantum dots in microcavities. iii. multiple
  quantum dots.
\newblock {\em Phys. Rev. B}, 84:195313, 2011.

\bibitem{Gerard.TopicsAP.2003}
J.-M. G\'{e}rard.
\newblock Solid-state cavity-quantum electrodynamics with self-assembled
  quantum dots.
\newblock {\em Topics Appl. Phys.}, 90:269--315, 2003.

\bibitem{Garcia.arxiv.2012}
P.~D. Garc\'{\i}a, S.~Stobbe, I.~S\"{o}llner, and P.~Lodahl.
\newblock Nonuniversal intensity correlations in 2d anderson localizing random
  medium.
\newblock arXiv:1206.3428, 2012.

\bibitem{Kaniber.PRB.2008}
M.~Kaniber, A.~Laucht, A.~Neumann, J.~M. Villas-B\^{o}as, M.~Bichler, M.-C.
  Amman, and J.~J. Finley.
\newblock Investigation of the nonresonant dot-cavity coupling in
  two-dimensional photonic crystal nanocavities.
\newblock {\em Phys. Rev. B}, 77:161303, 2008.

\bibitem{Diniz.PRA.2011}
I.~Diniz, S.~Portolan, R.~Ferreira, J.~M. G\'{e}rard, P.~Bertet, and
  A.~Auff\`{e}ves.
\newblock Strongly coupling a cavity to inhomogeneous ensembles of emitters:
  Potential for long-lived solid-state quantum memories.
\newblock {\em Phys. Rev. A}, 84:063810, 2011.

\bibitem{Stobbe.PRB.2010}
S.~Stobbe, T.~W. Schlereth, S.~H\"{o}fling, A.~Forchel, J.~M. Hvam, and
  P.~Lodahl.
\newblock Large quantum dots with small oscillator strength.
\newblock {\em Phys. Rev. B}, 82:233302, 2010.

\end{thebibliography}

\end{document}